\documentclass[12pt, letterpaper]{article}

\usepackage{arxiv}
\usepackage[utf8]{inputenc} 
\usepackage[T1]{fontenc}    
\usepackage{amsfonts}       

\usepackage{enumitem}
\usepackage{amssymb}
\usepackage{mathtools}
\usepackage[ruled,vlined,linesnumbered]{algorithm2e}
\usepackage{xcolor}
\usepackage{tikz}
\usepackage{pgfplots}
\usepackage{booktabs}
\usepackage[square,numbers]{natbib}
\usepackage{authblk}
\usepackage[toc, page]{appendix}
\usepackage[amsthm, amsmath, thmmarks]{ntheorem}

\allowdisplaybreaks[4]

\usetikzlibrary{
    patterns,
}
\theorempostskip{0pt}

\let\oldnl\nl
\newcommand{\nonl}{\renewcommand{\nl}{\let\nl\oldnl}}

\newtheorem{theorem}{Theorem}
\newtheorem{lemma}{Lemma}

\newtheorem{definition}{Definition}
\newtheorem{corollary}{Corollary}

\title{Optimal mechanisms with budget for user generated contents}

\author[1]{Mengjing Chen\thanks{ccchmj@qq.com}\enspace}
\author[1]{Pingzhong Tang\thanks{kenshinping@gmail.com}\enspace}
\author[2]{Zihe Wang\thanks{wang.zihe@mail.shufe.edu.cn}\enspace}
\author[1]{Shenke Xiao\thanks{xsk15@mails.tsinghua.edu.cn}\enspace}
\author[3]{Xiwang Yang\thanks{yangxiwang@bytedance.com}\enspace}
\affil[1]{Tsinghua University}
\affil[2]{Shanghai University of Finance and Economics}
\affil[3]{ByteDance}

\begin{document}
\maketitle
\vskip 1in
\begin{abstract}

In this paper, we design gross product maximization mechanisms which incentivize users to upload high-quality contents on user-generated-content (UGC) websites. We show that, the proportional division mechanism, which is widely used in practice, can perform arbitrarily bad in the worst case. The problem can be formulated using a linear program with bounded and increasing variables. We then present an $O(n\log n)$ algorithm to find the optimal mechanism, where $n$ is the number of players.
 
\end{abstract}

\newpage

\section{Introduction}\label{intro}

User-generated-content (UGC) websites \cite{luca2015user,zhang2014differentiation,park2014exploring} refer to those whose contents rely on users' post. 
In the last decades, we have witnessed an increasing number of UGC websites such as video-sharing website {\em YouTube}, question-and-answers website {\em Quora} and online encyclopedia {\em Wikipedia} where users generate contents autonomously.
On such websites, users are both consumers who view contents and contributors who post contents. High-quality contents play an important role in the success of these websites. Therefore a fundamental challenge faced by the UGC websites is to incentivize users to contribute high-quality contents. In this paper, we consider the problem of maximizing quality of all contents generated by users through a monetization reward mechanism \cite{ghosh2013learning,jain2014designing}.





In general, the center designs a reward function which maps a profile of users' contributed contents to a reward allocation. There are two natural models for such a reward function: a competitive model \cite{siegel2009all,chawla2012optimal,dipalantino2009crowdsourcing,ghosh2012crowdsourcing} or an independent model \cite{cavallo2012efficient,jain2013game}. For example, a reward function used by the Olympic games depends on the rank of contribution among all players and thus it is a competitive model. \citet{ghosh2014game} analyze the equilibrium in such rank-order mechanisms, where impressions are allocated in the decreasing order of content qualities. In particular, they also analyze the proportional mechanism, where impressions are allocated in proportional to qualities.  
\citet{luo2014optimal} give an optimal solution to the all-pay contests where each agent's type (ability to generate good contents) is private and is drawn from a known distribution. They allow personalized reward function that can be different for different agents, while we impose the restriction of anonymity which is required by almost all these UGC sites.

This paper focuses on the reward function for the independent model where an agent's reward doesn't depend on the quality other agents contribute. For example, a taxation rule used by government depends on how much a person earned regardless of other people's status \cite{mirrlees1971exploration,sheshinski1972optimal}. The merit of such an independent model is that it is easier for a user to compute the best strategy, and thus it is widely adopted in large-scale practical markets.

As a result, we restrict the design space to a universal reward function which maps from a user's contribution to a non-negative real reward. \citet{dasgupta1998designing} considers a similar problem but they only focus on the case where the types of all agents are the same. In our model, we consider the general case where agents could have different types.

The goal of the center is to design such a reward mechanism to incentivize users to generate high-quality contents as much as possible. When a user generates a content, the user also takes a cost that depends on his type. Given the reward function, a user can choose how much effort to put in generating the content and will take the best action to maximize utility. We now present the center question studied in this paper:

If the website has a fixed budget, what is the optimal reward mechanism? 

Some mechanisms such as top-$K$ allocation, proportional allocation have been proven simple to reach Nash Equilibria among users. However, 
we will show that the proportional allocation can benefit the center as little as an $\epsilon$ fraction of the optimal objective.

\subsection{Our Techniques}
We first give a characterization of the optimal reward mechanism and formalize it as a linear program with variables $x_1^*,x_2^*,\ldots,x_n^*$ of the form
\begin{equation} \label{main form}
\begin{aligned}
\text{maximize} &&& \displaystyle\sum_{i=1}^n x_i^* \\
\text{subject to} &&& 0\le x_i^* \le q_i, & i=1,2,\ldots,n,\\
                        &&& 0\le x_1^*\le x_2^*\le\cdots\le x_n^*, \\
                        &&& \displaystyle \sum_{i=1}^nz_ix_i^*\le K,
\end{aligned}
\end{equation}
where $0< q_1\le \cdots \le q_n$, $z_i> 0$ and $K> 0$. The connection between these variables and the mechanism will be discussed in detail in the context. Here we focus on this linear program. 

Although there are various techniques of solving standard linear programmings \cite{dantzig2006linear}, linear programmings with specific forms are also studied in the literature. \citet{andersson2006fast} study a similar linear programming where each constraint has the form $x_i\ge\lambda x_j+\beta$ for some $\beta,\lambda\in\mathbb{R}$ with $0<\lambda<1$, and each variable appears in the left-hand side of at least one constraint. \citet{burkard2008inverse} study a specific kind of linear programming with bounded constraints $0\le x_i\le \bar{x}_i$. However, their techniques do not help to solve our Problem (\ref{main form}).

In Section \ref{A Greedy Algorithm}, we propose a greedy algorithm, which is, to our knowledge, the first efficient algorithm that solves linear programmings of the form (\ref{main form}). 
This algorithm initializes $x_1^*=x_2^*=\cdots=x_{i-1}^*=0$, and maintains a set $S$ of indices that indicate which $x_i^*$'s are currently full (i.e. $x_i^*=q_i$).
The set $S$ is initialized with $\{0,n+1\}$ and the algorithm iteratively adds an index $i$ to $S$ in a greedy principle such that $i$ minimizes the average value of $z_i,z_{i+1},\ldots,z_{\min_{j>i:j\in S}j-1}$. At the same time it increases simultaneously $x_i^*,x_{i+1}^*,\ldots,x_{\min_{j>i:j\in S}j-1}^*$ by a value as large as possible, which means either $x_i^*$ reaches the bound $q_i$ or $\sum_{i=1}^nz_ix_i^*$ reaches the bound $K$. 
This algorithm runs in $\Theta(n^2)$ time because one needs to maintain the average value of $z_i,z_{i+1},\ldots,z_{\min_{j>i:j\in S}j-1}$ for each $i$, and each time an index is added to $S$, it takes $\Theta(n)$ time to update these average values. Observing that for any $i$, the average value of $z_i,z_{i+1},\ldots,z_{\min_{j>i:j\in S}j-1}$ is only used in the iteration where $i$ is chosen (let us denote by $\mathrm{avg}_i$ this average value)\footnote[1]{This average value varies as $S$ grows in each iteration, but we are only interested in its value in the iteration where $i$ is chosen, which is denoted by $\mathrm{avg}_i$.}, it is straightforward to consider computing $\mathrm{avg}_i$ in advance instead of maintaining the average value as $S$ grows.
In Section \ref{A More Efficient Algorithm}, we prove that if for each $i$, we compute $\mathrm{avg}_i$ at the very beginning of the algorithm, and does not update it, the algorithm is still correct. The improved algorithm runs in $O(n\log n)$ time.
Although our problem is motivated in the context of UGC website, our algorithm is of independent interest for solving linear programs of the form (\ref{main form}). 



%

\section{Preliminaries} \label{Preliminaries}
Let $N=(1,2,\dots, n)$ be the set of all agents in a UGC website. Each agent $i$ has a private type $q_i \in \mathbb{R}_+$, which stands for the best quality of content that he can produce. Without loss of generally, we assume $0 < q_1 \leq q_2 \leq \dots \leq q_n$. In this paper, we analyze the problem of incentivizing high-quality contents in the {\em full information setting} where the type profile $(q_1,q_2,\ldots,q_n)$ is known to all agents and the website. The action of each agent $i$ is posting a content with quality $x_i$ on the website where $x_i$ cannot exceed her type $q_i$. In this paper, we consider a continuous action space, namely, $x_i \in [0, q_i]$. 
The cost for agent $i$ to produce a content with quality $x_i$ is $c_i = x_iC/q_i$, where $C$ is a positive constant. 

Given a fixed budget $B$, the website aims to design a reward mechanism that maximizes the gross product, which is defined as the overall quality of all contents on the website, i.e., $\sum_{i \in N} x_i$. The reward mechanism specifies each agent's reward when the budget and agents' types are given. Formally,

\begin{definition}[Reward mechanism] A reward mechanism is a reward function $f$ where
$f: \mathbb{R}+ \mapsto \mathbb{R}+$ is the mapping from the quality of a content to the reward. 
\end{definition}

Note that the reward function only takes one agent's action as the input, which means the reward an agent receives is only based on the quality of content he produces and independent of other agents' actions. This mechanism is simple and easily understood by agents, also makes agents pay more attention to their own contents instead of the environment.

We assume that agents can get no utilities except for the reward on the website, thus the utility of agent $i$ is the reward she receives minus the cost of producing the content, i.e., 
\begin{gather*} 
u_i(x_i) = f(x_i) - \frac{x_i C}{q_i}. 
\end{gather*}
All agents are strategic, meaning that they will give the best responses, choosing qualities of contents which maximize their utilities, to the reward function.

With strategic agents, our goal is to find an optimal reward function $f$ that results in the maximal gross product. The problem we describe above can be represented as the optimization problem presented as below. 
\begin{align}\label{free problem}
\begin{aligned}
\text{maximize}  &&& \displaystyle \sum_{i=1}^n x_i^* \\
\text{subject to} &&& \text{(utility definition)} && u_i(x_i)=f(x_i)-\frac{x_iC}{q_i} \\ &&& 
     \text{(incentive constraint)} &&  \forall x\in [0,q_i], u_i(x_i^*)\ge u_i(x),  && i=1,2,\ldots,n \\
                          &&&  \text{(budget constraint)} &&  \displaystyle\sum_{i=1}^n f(x_i^*)\le B\\
                          &&&  \text{(capability constraint)}  &&  0\le x_i^* \le q_i, && i=1,2,\ldots,n\\
                          &&& \text{(non-negative reward)} &&  \forall x\ge 0, f(x)\ge 0
\end{aligned}
\end{align}

\section{The Proportional Mechanism} \label{The Proportional Mechanism}
In this section, we introduce the widely used mechanism, the proportional mechanism. In the proportional mechanism, agents share the total reward in proportional to the qualities of contents they produce. Formally, 
the utility of agent $i$ in this mechanism can be represented as
\[u_i(x_i,x_{-i})=\dfrac{x_iB}{\sum_{j=1}^nx_j}-\dfrac{x_iC}{q_i}.\]
For completeness, let $u_i(x_i,x_{-i})$ be $0$ if $x_i=0$ for all $i$.

In this mechanism, the utility of each agent depends not only on the quality of the content she produces, but also on the qualities of the contents other agents produce. However, this mechanism is very inefficient, in the sense that the ratio of the gross product of the website under any Nash equilibrium to that under any optimal solution in our mechanism can be infinitely small. We state its inefficiency on gross product  in the following theorem. All the missing proofs in this paper are deferred to the appendices.

\begin{theorem} \label{prop theorem}
Even if there are 2 agents and the total budget $B$ equals to the cost constant $C$, for any $\epsilon>0$, there exists a type profile $(q_1,q_2)$ of the agents such that for any pure Nash equilibrium $(x_1^{\text{prop}}, x_2^{\text{prop}})$ of the proportional mechanism and for any optimal solution $(x_1^*, x_2^*)$ to Problem (\ref{free problem}), we have
\[\frac{x_1^{\text{prop}}+x_2^{\text{prop}}}{x_1^*+x_2^*}\le \epsilon.\]
\end{theorem}

This theorem shows that our mechanism corresponding to Problem (\ref{free problem}) beats the proportional mechanism a lot.

\section{The optimal Mechanism} \label{Equivalent Linear Programming}

The problem we formulate in Section \ref{Preliminaries} is complicated since the optimization variable is a mapping, which has a huge design space. In this section, we prove that the original problem can be solved in polynomial time. We first show that there always exists an optimal solution such that $x_1^*\le x_2^*\le\cdots \le x_n^*$, which implies an agent with higher type will post a content with higher quality. Then we characterize the optimal piecewise reward function. By taking advantages of such characterization, we formulate a linear programming to find the optimal solution.

\begin{lemma}\label{free model x lemma}
For any feasible solution $(f,x_1^*,x_2^*,\ldots,x_n^*)$ to Problem (\ref{free problem}), there exists a feasible solution $(f,x_1',x_2',\ldots,x_n')$ to (\ref{free problem}) such that $x_1'\le x_2'\le\cdots\le x_n'$ and $\sum_{i=1}^n x_i'\ge\sum_{i=1}^n x_i^*$.
\end{lemma}

Lemma \ref{free model x lemma} implies that we can only focus on the ordinal strategy profile $(x_1^*,x_2^*,\ldots,x_n^*)$ where $0\le x_i^*\le q_i$ for each $i$, and $x_1^*\le x_2^*\le\cdots\le x_n^*$. However, this characterization is not enough to make $x_1^*,x_2^*,\ldots,x_n^*$ feasible because there may not exist a reward function $f$ that incentivizes the agents. Given the ordinal strategy profile $(x_1^*,x_2^*,\ldots,x_n^*)$, the following lemma gives a necessary and sufficient condition of $f$ that makes $(f,x_1^*,x_2^*,\ldots,x_n^*)$ be a feasible solution of Problem (\ref{free problem}).

\begin{lemma} \label{free model f lemma}
Given $x_1^*,x_2^*,\ldots,x_n^*$ such that $0\le x_i^*\le q_i$ for each $i$, and $x_1^*\le x_2^*\le\cdots\le x_n^*$, there exists a function $f$ such that $(f,x_1^*,\ldots,x_n^*)$ is a feasible solution to (\ref{free problem}) if and only if
\begin{equation}\label{free model budget constraint}
C\left(\frac{x_n^*}{q_n}+\sum_{i=1}^{n-1}\left((n-i)\left(\frac{1}{q_i}-\frac{1}{q_{i+1}}\right)+\frac{1}{q_i}\right)x_i^*\right)\le B.
\end{equation}
Moreover, if (\ref{free model budget constraint}) is satisfied, to make $(f,x_1^*,\ldots,x_n^*)$ a feasible solution, we can set 
\begin{equation} \label{free f}
f(x)=\begin{dcases}
0, & \text{if } 0\le x<x_1^*, \\
\dfrac{x_1^*C}{q_1}, & \text{if } x_1^*\le x<x_2^*, \\
\left(\dfrac{x_1^*}{q_1}-\dfrac{x_1^*}{q_2}+\dfrac{x_2^*}{q_2}\right)C, & \text{if } x_2^*\le x<x_3^*, \\
\cdots, \\
\left(\dfrac{x_1^*}{q_1}-\dfrac{x_1^*}{q_2}+\dfrac{x_2^*}{q_2}-\dfrac{x_2^*}{q_3}+\cdots+\dfrac{x_n^*}{q_n}\right)C, & \text{if } x\ge x_n^*.
\end{dcases}
\end{equation}
\end{lemma}

The two lemmas above induce an equivalent linear programming to solve the original problem.
\begin{theorem}
Problem (\ref{free problem}) has the same optimal value as the following linear programming:
\begin{equation}\label{free linear problem}
\begin{aligned}
\text{maximize} &&& \displaystyle\sum_{i=1}^n x_i^* \\
\text{subject to} &&& 0\le x_i^* \le q_i, & i=1,2,\ldots,n,\\
                          &&& 0\le x_1^*\le x_2^*\le\cdots\le x_n^*, \\
                          &&& \displaystyle C\left(\frac{x_n^*}{q_n}+\sum_{i=1}^{n-1}\left((n-i)\left(\frac{1}{q_i}-\frac{1}{q_{i+1}}\right)+\frac{1}{q_i}\right)x_i^*\right)\le B.
\end{aligned}
\end{equation}
Once an optimal solution $(x_1^*,x_2^*,\ldots,x_n^*)$ is obtained, an optimal reward function can be constructed as (\ref{free f}).
\end{theorem}

For ease of representation, let $z_i$ be the coefficient of $x_i^*$ in the budget constraint in Problem (\ref{free linear problem}), that is, $z_i=(n-i)(1/q_i-1/q_{i+1})+1/q_i$ for $i=1,2,\ldots, n-1$, and $z_n=1/q_n$, then we can see that Problem (\ref{free linear problem}) is exactly the form of Problem (\ref{main form}), where $K=B/C$. In the rest of the paper, we will focus on Problem (\ref{main form}).

\section{A Greedy Algorithm} \label{A Greedy Algorithm}

In this section, we propose a $\Theta(n^2)$ greedy algorithm that solves Problem (\ref{main form}) (thus Problem (\ref{free linear problem}) is also solved). We first present our algorithm and then show the output of the algorithm is exactly an optimal solution of Problem (\ref{main form}). 

For ease of representation, let us define
\[\mathrm{sum}(i,j)=z_i+z_{i+1}+\cdots+z_{j-1},\quad \mathrm{avg}(i,j)=\frac{\mathrm{sum}(i,j)}{j-i}.\]
In Problem (\ref{free linear problem}), since for any $i<j\le n$, $\mathrm{sum}(i,j)=(n-i+1)/q_i-(n-j+1)/q_j$, and $\mathrm{sum}(i,n+1)=(n-i+1)/q_i$, both the functions $\mathrm{sum}$ and $\mathrm{avg}$ can be computed in $O(1)$ time. In the more general Problem (\ref{main form}), we still assume that they can be computed in $O(1)$ time. For the $\Theta(n^2)$ algorithm proposed in this section, this assumption is reasonable because we can compute $\mathrm{sum}(i,j)$ and $\mathrm{avg}(i,j)$ for each $i,j$ in advance, which does not increase the time complexity of the algorithm. We will give further explanation why this assumption is reasonable at the end of Section \ref{A More Efficient Algorithm}.

Note there are three kinds of constraints in Problem (\ref{main form}), and it is their combination that makes this problem non-trivial:
\begin{enumerate}
\item Without the constraints $x_i^*\le q_i$, it is optimal to set $x_1^*=x_2^*=\cdots=x_{i-1}^*=0$ and $x_{i-1}^*=x_{i-2}^*=\cdots=x_n^*=K/\mathrm{sum}(i,n+1)$ where $i$ is the index that minimizes $\mathrm{avg}(i,n+1)$.
\item Without the constraint $0\le x_1^*\le\cdots\le x_n^*$, the following simple algorithm would output an optimal solution: initializing $x_1^*=x_2^*=\cdots=x_n^*=0$, and then increasing $x_i^*$ to $q_i$ for each $i$ in the order for $z_i$'s from large to small, until $\sum_{i=1}^n z_ix_i^*$ reaches $K$.
\item Without the constraint $\sum_{i=1}^n z_ix_i^*\le K$, it is optimal to set $x_i^*=q_i$ for each $i$.
\end{enumerate}
When all the three kinds of constraints exist, none of the methods above applies, which makes Problem (\ref{main form}) hard to solve.

Our algorithm can be considered as a combination of the three methods above. In short, our algorithm initializes $x_1^*=x_2^*=\cdots=x_{i-1}^*=0$, and maintains a set $S$ of indices that indicate which $x_i^*$'s are currently full (i.e., $x_i^*=q_i$). The set $S$ is initialized with $\{0,n+1\}$ and the algorithm iteratively adds an index $i$ to $S$ in a greedy principle such that $i$ minimizes the average value of $z_i,z_{i+1},\ldots,z_{\min_{j>i:j\in S}j-1}$. At the same time it increases simultaneously $x_i^*,x_{i+1}^*,\ldots,x_{\min_{j>i:j\in S}j-1}^*$ by a value as large as possible, which means either $x_i^*$ reaches the bound $q_i$ or $\sum_{i=1}^nz_ix_i^*$ reaches the bound $K$. This whole algorithm is formally shown as Algorithm \ref{free model algorithm}. 

Suppose Algorithm \ref{free model algorithm} runs for $k$ iterations in total, and Line \ref{free model algorithm line loop begin} chooses $i^*$ to be $i_1,i_2,\ldots,i_k$ in order. We define $S_\ell$ to be the set $S$ after the $\ell$th iteration, i.e., $S_\ell=\{0,i_1,\ldots,i_\ell,n+1\}$. In addition, we define
\[\mathrm{left}_\ell(i)=\max_{j\in S_\ell:j<i}j,\quad \mathrm{right}_\ell(i)=\min_{j\in S_\ell:j>i}j.\]
Note by these definitions, we have in the $\ell$th iteration, $i_L=\mathrm{left}_\ell(i^*)$ and $i_R=\mathrm{right}_\ell(i^*)$. 

The notations used to analyze Algorithm \ref{free model algorithm} are summarized in Table \ref{notations table}.

Before we prove the correctness of this algorithm, we state Lemma \ref{xyB lemma} to show how the values of $x_i,y_i$'s and $\hat{B}$ change in the algorithm.
\begin{table}[htbp!]
\centering
\caption{Notations} \label{notations table}
\begin{tabular}{ll}
\toprule
Notations & Meanings\\
\midrule
$n,q_i,K$ & the parameters in Problem (\ref{main form}) \\
$x_1^*,x_2^*,\ldots,x_n^*$ & the variables in Problem (\ref{main form}), usually mentioned along with an optimal solution\\
$i_L,i_R,d,x_i,y_i,\hat{B},S$ & the variables used in the description of Algorithm \ref{free model algorithm}\\
$k$ & the number of iterations for which Algorithm \ref{free model algorithm} runs in total \\
$i_1,i_2,\ldots,i_k$ & Line \ref{free model algorithm line loop begin} of Algorithm \ref{free model algorithm} chooses $i^*$ to be $i_1,i_2,\ldots,i_k$ in order  \\
$S_\ell$ & the set $S$ immediately after the $\ell$th iteration \\
$\mathrm{left}_\ell(i)$ & $\max_{j\in S_\ell:j<i}j$ \\
$\mathrm{right}_\ell(i)$ & $\min_{j\in S_\ell:j>i}j$ \\
\bottomrule
\end{tabular}
\end{table}

\begin{figure}[htbp!]
\begin{minipage}[t]{0.47\linewidth}
\vspace{0pt}
\centering
\begin{algorithm}[H] 
$S\leftarrow\{0,n+1\}$\;
\For {$i\leftarrow1$ \KwTo $n$} {
 $\displaystyle y_i\leftarrow \mathrm{avg}(i,n+1)$\;
 $x_i\leftarrow 0$\;
}
$\hat{B}\leftarrow K$\;
\While {$\hat{B}>0$ and $S\neq \{0,1,\ldots,n+1\}$} {
    $\displaystyle i^*\leftarrow \arg\min_{i\notin S}(y_i, i)$\tcp*{$(a,b)<(c,d)$ if and only if $a<c$, or $a=c$ and $b<d$, so if there are minimum values, the algorithm will choose the one with the smallest index} \label{free model algorithm line loop begin}
    $\displaystyle i_L\leftarrow\max_{i\in S:i<i^*}i$\;
    $\displaystyle i_R\leftarrow\min_{i\in S:i>i^*}i$\; 
    $\displaystyle d \leftarrow\min\left\{\frac{\hat{B}}{(i_R-i^*)y_{i^*}}, q_{i^*}-x_{i^*}\right\}$\; \label{free model algorithm line compute d}
    $\hat{B} \leftarrow \hat{B}- d(i_R-i^*)y_{i^*}$\; \label{free model algorithm line update B}
    \For {$i\leftarrow i^*$ \KwTo $i_R-1$} {
      $\displaystyle x_i\leftarrow x_i+d$\; \label{free model algorithm line update x}
    }
    \For {$i\leftarrow i_L+1$ \KwTo $i^*-1$} { \label{free model algorithm line update y loop}
      $\displaystyle y_i\leftarrow \mathrm{avg}(i,i^*)$\;  \label{free model algorithm line update y}
    }
    add $i^*$ to $S$\; \label{free model algorithm line update S}
}
output $x_1,\ldots,x_n$ as $x_1^*,\ldots,x_n^*$\;
\caption{An $O(n^2)$ Algorithm to Solve Problem (\ref{main form})} \label{free model algorithm}
\end{algorithm}
\end{minipage}
\hspace{0.06\linewidth}
\begin{minipage}[t]{0.47\linewidth}
\vspace{0pt}
\centering
\begin{algorithm}[H] 
$S\leftarrow\{0,n+1\}$\;
$\displaystyle b_n\leftarrow n+1$\tcp*{$b_i$ is used to compute $b(i)$}
$\displaystyle y_n\leftarrow \frac{1}{q_n}$\;
\For {$i\leftarrow n-1$ \KwTo $1$} {
 $b_i\leftarrow i + 1$\;
 \While {$b_i\neq n+1$ and $\mathrm{avg}(i,b_i)\le\mathrm{avg}(b_i, b_{b_i})$} {\label{improved free model algorithm line update b loop}
     $b_i\leftarrow b_{b_i}$\; \label{improved free model algorithm line update b}
 }
 $\displaystyle y_i\leftarrow \mathrm{avg}(i,b_i)$\; \label{improved free model algorithm line update y}
 $x_i\leftarrow 0$\;
}
$\hat{B}\leftarrow K$\;
\While {$\hat{B}>0$ and $S\neq \{0,1,\ldots,n+1\}$} {
    $\displaystyle i^*\leftarrow \arg\min_{i\notin S}(y_i, i)$\tcp*{$(a,b)<(c,d)$ if and only if $a<c$, or $a=c$ and $b<d$, so if there are minimum values, the algorithm will choose the one with the smallest index}
    $\displaystyle d \leftarrow\min\left\{\frac{\hat{B}}{(b_i-i^*)y_{i^*}}, q_{i^*}-x_{i^*}\right\}$\;
    $\hat{B} \leftarrow \hat{B}- d(b_i-i^*)y_{i^*}$\;
    \For {$i\leftarrow i^*$ \KwTo $b_i-1$} { \label{improved free model algorithm line update x loop}
      $\displaystyle x_i\leftarrow x_i+d$\; \label{improved free model algorithm line update x}
    }
    add $i^*$ to $S$\;
}
output $x_1,\ldots,x_n$ as $x_1^*,\ldots,x_n^*$\;
\caption{Improved Algorithm \ref{free model algorithm}} \label{improved free model algorithm}
\end{algorithm}
\end{minipage}
\end{figure}

\begin{lemma} \label{xyB lemma}
Immediately after the $\ell$th ($\ell< k$) iteration, for each $i$, we have
\begin{align}
x_i&=
\begin{cases}
q_i, &\text{if $i\in S_\ell$,}\\  
x_{i-1}, &\text{otherwise,}
\end{cases} \label{x} \\
y_i&=\mathrm{avg}\left(i,\mathrm{right}_\ell(i)\right), \label{y} \\
\hat{B}&=K-\sum_{i=1}^nz_ix_i. \label{B}
\end{align}
These properties also hold for $\ell=k$ except that $x_{i_{k}}$ is not necessarily equal to $q_{i_k}$.
\end{lemma}

This lemma can be proven by observing that these equations hold initially, and the updates for $x_i,y_i$ and $S$ in each iteration of Algorithm \ref{free model algorithm} do not invalidate them. We omit the detail of this proof.

Lemma \ref{B lemma}  is a supplement of Lemma \ref{xyB lemma}. It shows what the values of $x_i$'s are after the last iteration.

\begin{lemma} \label{B lemma}
After the last iteration, either $\sum_{i=1}^n z_ix_i=K$, or $x_i=q_i$ for each $i$.
\end{lemma}

Now we begin to prove the correctness of Algorithm \ref{free model algorithm}. We first show that the subset $\{x_{i_1},x_{i_2},\ldots,x_{i_k}\}$ of the output of Algorithm \ref{free model algorithm} matches an optimal solution.

\begin{lemma} \label{free model algorithm correctness lemma}
Let $x_1,x_2,\ldots,x_n$ be the variables after the last iteration. There exists an optimal solution for Problem (\ref{main form}) such that $x_{i_1}^*=x_{i_1},x_{i_2}^*=x_{i_2},\ldots$, and $x_{i_k}^*=x_{i_k}$.
\end{lemma}

With Lemma \ref{free model algorithm correctness lemma}, it is not hard to prove the correctness of Algorithm \ref{free model algorithm}. We state it as Theorem \ref{correctness theorem}.

\begin{theorem} \label{correctness theorem}
The optimal solution to Problem (\ref{main form}) can be solved by Algorithm \ref{free model algorithm}. 
\end{theorem}
\begin{proof}
Consider the optimal solution in Lemma \ref{free model algorithm correctness lemma}. Let $x_1,x_2,\ldots,x_n$ be the variables after the last iteration. By Lemma \ref{free model algorithm correctness lemma}, we have $x_i^*=x_i$ for all $x\in S_k$. Now let us fix some $i\notin S_k$. By the feasibility constraints in Problem (\ref{main form}), we have $x_i^*\ge x_{\mathrm{left}_k(i)}^*$. By Lemma \ref{free model algorithm correctness lemma}, we have $x_{\mathrm{left}_k(i)}^*=x_{\mathrm{left}_k(i)}$. By (\ref{x}) we have $x_{\mathrm{left}_k(i)}=x_i$. Hence we can conclude that for any $i$, no matter whether $i\in S_k$ or not, $x_i^*\ge x_i$. If there exists some $i$ such that $x_i^*>x_i$, then $x_i<q_i$, and by Lemma \ref{B lemma}, $\sum_{i=1}^n z_ix_i=K$, so $\sum_{i=1}^n z_ix_i^*>\sum_{i=1}^n z_ix_i=K$, which contradicts to the feasibility constraints in Problem (\ref{main form}). As a result, for any $i$, $x_i^*=x_i$. This algorithm is an optimal algorithm.
\end{proof}

\section{A More Efficient Algorithm} \label{A More Efficient Algorithm}

Note in Algorithm \ref{free model algorithm}, the variables $x_i$'s and $y_i$'s are updated many times by Line \ref{free model algorithm line update x} and \ref{free model algorithm line update y} in each iteration, which are two main bottlenecks that make this algorithm run in $\Theta(n^2)$ time. In this section, we aim to improve this algorithm by avoiding redundant computation of $x_i$'s and $y_i$'s. The improved algorithm runs in $O(n\log n)$ time.

We first solve the bottleneck for $y_i$'s. For any $i$, let us denote by $b(i)$ (the \emph{blocker} for $i$) the final $i^*$ in the last update for $y_i$. If there is no blocker for $i$, i.e., $y_i$ keeps $1/q_i$ all along, let $b(i)=n+1$ for convenience. In addition, let $b^0(\cdot)$ be the identity function, and $b^m(\cdot)=b(b^{m-1}(\cdot))$. The following lemma shows that if we know $b(i)$ for each $i$ in advance, we can improve our algorithm by directly assigning $\mathrm{avg}(i,b(i))$ to $y_i$ at the beginning, no need of updating $y_i$ any more.

\begin{lemma}\label{yi lemma}
If we modify Algorithm \ref{free model algorithm} by initializing $y_i$ to be $\mathrm{avg}(i,b(i))$ (before the \textbf{while} loop) and not updating them any more (deleting Line \ref{free model algorithm line update y loop} to \ref{free model algorithm line update y}), then for any inputs, the modified version has the same outputs as Algorithm \ref{free model algorithm}'s.
\end{lemma}

Lemma \ref{yi lemma} gives us insights to design more efficient algorithms. If the blockers are known in advance, a faster algorithm gets naturally. Hence the main challenge becomes finding the blocker for each $i$. We will state some properties of the blockers first in the following lemmas. Then we will show how to design an algorithm to get such blockers by using their properties. 


\begin{lemma} \label{iRb lemma}
In each iteration, $i_R=b(i^*)$, which means all indices between $i^*$ (including) and $b(i^*)$ (excluding) do not belong to $S$ and $b(i^*)\in S$ for the set $S$ immediately before this iteration.
\end{lemma}

Lemma \ref{order lemma} shows the execution order of iterations where $i^*$ is chosen to be $i,b(i)$ and the indices between them.

\begin{lemma} \label{order lemma}
For any $i$, if $b(i)\le n$, then 
\begin{enumerate}[label=\roman*)]
\item the iteration where $i^*=i$ comes after the iteration where $i^*=b(i)$, and
\item for any $i<j<b(i)$, the iteration where $i^*=j$ comes after the iteration where $i^*=i$.
\end{enumerate}
\end{lemma}

Lemma \ref{avg lemma} shows two inequality relations among $\mathrm{avg}$'s related to $i_L,i^*,i_R$ in each iteration. Figure \ref{inequalities figure} diagrams these relations.
\begin{lemma} \label{avg lemma}
In each iteration, for any $i^*<i<b(i^*)$, we have
$
\mathrm{avg}(i^*,i)\le \mathrm{avg}(i,b(i^*))
$, 
and for any $i_L<i<i^*$, we have $\mathrm{avg}(i,b(i^*))<\mathrm{avg}(i, i^*)$.
\end{lemma}

\begin{figure}
\centering
\begin{tikzpicture}[scale=1]
\draw [thick]  (0,0) -- (0.5,0);
\draw [thick, dashed]  (0.5,0) -- (1.5,0);
\draw [thick]  (1.5,0) -- (8.5,0);
\draw [thick, dashed]  (8.5,0) -- (9.5, 0);
\draw [thick]  (9.5, 0) -- (10,0);
\draw (0,-.2) -- (0, .2);
\draw (10,-.2) -- (10, .2);
\draw (2,-.2) -- (2, .2);
\draw (5,-.2) -- (5, .2);
\draw (8,-.2) -- (8, .2);
\draw[pattern=north east lines] (5,0.5) rectangle (6,0.7);
\draw[pattern=north east lines] (6.1,0.5) rectangle (7.9,0.7);
\draw[pattern=north east lines] (3,-1.9) rectangle (7.9,-1.7);
\draw[pattern=north east lines] (3,-2.4) rectangle (4.9,-2.2);
\node[align=left, below] at (0,-.3)
    {$0$};
\node[align=left, below] at (2,-.3)
    {$i_L$};
\node[align=left, below] at (5,-.3)
    {$i^*$}; 
\node[align=left, below] at (8,-.3)
    {$i_R$};
\node[align=left, below] at (0,-1.8)
    {$\mathrm{avg}(i,b(i^*))<\mathrm{avg}(i, i^*)$};
\draw[->] (-1, -1.75) to [out=30,in=160] (2.8, -1.4);
\draw[->] (1.2, -2.5) to [out=-30,in=200] (2.8, -2.7);
   
\node[align=left, below] at (6.3,1.5)
    {$\mathrm{avg}(i^*,i)\le \mathrm{avg}(i,b(i^*))$};
\node[align=left, below] at (8,-0.85)
    {$b(i^*)$};  
\node[align=left, below] at (10,-.3)
    {$n+1$};
\end{tikzpicture}
\caption{$\mathrm{avg}$ Inequalites} \label{inequalities figure}
\end{figure}

\begin{lemma} \label{iL lemma}
In any iteration, if there exists some $m$ and $j$ such that $i^*=b^m(j)$, then $i_L<j$.
\end{lemma}

By combining Lemma \ref{avg lemma} and \ref{iL lemma}, we can get the following corollary immediately.

\begin{corollary} \label{avg inequality2 corollary}
In any iteration, if there exists some $m\ge 1$ and $j$ such that $i^*=b^m(j)$, then
$
\mathrm{avg}(j,b(i^*))<\mathrm{avg}(j, i^*).
$
\end{corollary}

With the properties stated above, we can show the following important property of $b(i)$, which suggests us a way to find $b(i)$ for each $i$.

\begin{theorem}\label{improved free model algorithm lemma}
For any $i$, there exists some $m$ such that $b(i)=b^m(i+1)$. Moreover, for any $0\le t<m$, we have
\begin{equation} \label{first main inequality}
\mathrm{avg}(i,b^t(i+1))\le\mathrm{avg}(b^t(i+1), b^{t+1}(i+1)),
\end{equation}
and if $b^{m+1}(i+1)$ exists,
\begin{equation} \label{second main inequality}
\mathrm{avg}(i,b^m(i+1))>\mathrm{avg}(b^m(i+1), b^{m+1}(i+1)).
\end{equation}
\end{theorem}
\begin{proof}
We first show that $b(i)=b^m(i+1)$ for some $m$. We can assume there exists $j$ such that $i<j<b(i)$, otherwise $b(i)=i+1=b^0(i+1)$. By Lemma \ref{order lemma}, the iteration where $i^*=b(i)$ comes before the iteration where $i^*=j$, so in the iteration where $i^*=j$, we have already $b(i)\in S$, thus $b(j)=i_R\le b(i)$. By choosing $j$ to be $i+1$, we have $b(i)\ge b(i+1)$. If $b(i)>b(i+1)$, we then choose $j$ to be $b(i+1)$, and have $b(i)\ge b^2(i+1)$. By applying this process repeatedly, we will have $b(i)=b^m(i+1)$ for some $m$ finally. 

Now let us fix a $t$ such that $0\le t<m$. For any $t< \tau< m$, consider the iteration where $i^*=b^{\tau}(i+1)$. By Corollary \ref{avg inequality2 corollary} we have $\mathrm{avg}(b^t(i+1),b^{\tau+1}(i+1))<\mathrm{avg}(b^t(i+1), b^{\tau}(i+1))$. By choosing $\tau$ to be $t+1,t+2,\ldots,m-1$ and combining these inequalities, we have $\mathrm{avg}(b^t(i+1),b^m(i+1))<\mathrm{avg}(b^t(i+1), b^{t+1}(i+1))$. Then we consider the iteration where $i^*=i$. By 
Lemma \ref{avg lemma} we have $\mathrm{avg}(i,b^t(i+1))\le\mathrm{avg}(b^t(i+1), b(i))$. Combining the two inequalities above, we have proved (\ref{first main inequality}). Moreover, if $b^{m+1}(i+1)$ exists, consider the iteration where $i^*=b^m(i+1)=b(i)$, then (\ref{second main inequality}) is implied by Corollary \ref{avg inequality2 corollary}.
\end{proof}

By Theorem \ref{improved free model algorithm lemma}, we can find $b(i)$ by checking whether $\mathrm{avg}(i,b^t(i+1))\le\mathrm{avg}(b^t(i+1), b^{t+1}(i+1))$ for $t=0,1,\ldots$, which results in Algorithm \ref{improved free model algorithm}. Now we have solved the bottleneck for $y_i$'s. The bottleneck for $x_i$'s is solved in the proof of Theorem \ref{nlogn theorem}.

\begin{theorem} \label{nlogn theorem}
Algorithm \ref{improved free model algorithm} runs in $O(n\log n)$ time.
\end{theorem}
\begin{proof}
The only tricky parts are Line \ref{improved free model algorithm line update b loop} to \ref{improved free model algorithm line update b} and Line \ref{improved free model algorithm line update x loop} to \ref{improved free model algorithm line update x}. 

Now let us fix an index $j$. Suppose Line 6 accesses $j$ as $b_i$ and the condition in the while loop still holds, which means $i<j<b(i)$. We can prove by mathematical induction that for any $i'<i$, there does not exist $m$ such that $b^m(i'+1)=j$, which means $j$ will not be accessed by Line 6 any more. In other words, except for the last iteration in the loop corresponding to to Line 6 to 7, each index is accessed as $b_i$ at most once by Line 6 during the whole algorithm. Hence Line 4 to 9 cost $O(n)$ time in total. 

To efficiently update $x_i$'s in Line \ref{improved free model algorithm line update x loop} to \ref{improved free model algorithm line update x}, we can use a binary indexed tree. More precisely, we maintain an array $A$ of size $n$. Initially $A$ is filled with zeros. Each time we add $d$ to $x_i,x_{i+1}\ldots,x_j$, we instead increase $A[i]$ by $d$ and if $j<n$, decrease $A[j+1]$ by $d$. Each time we access $x_i$, we instead return the initial value of $x_i$ plus $A[1]+A[2]+\cdots+A[i]$. With a binary indexed tree, both operations can be done in $O(\log n)$ time. Hence updating and accessing $x_i$'s in the whole algorithm cost $O(n\log n)$ time.

So the whole algorithm runs in $O(n\log n)$ time.
\end{proof}

Recall that the analysis above is based on the fact that $\mathrm{avg}(i,j)$ can be computed in $O(1)$ time for any $i<j$. This is true for our UGC website problem, but what if we are facing a general problem of the form (\ref{main form}), where $\mathrm{avg}(i,j)$ cannot be computed in $O(1)$ time? Is Algorithm \ref{improved free model algorithm} still able to run in $O(n\log n)$ time? 

Note in Line \ref{improved free model algorithm line update b loop} to \ref{improved free model algorithm line update y}, since $b_i>i$, the initialization for $y_{b_i}$ is completed, i.e., the value of $\mathrm{avg}(b_i, b_{b_i})$ is already recorded in $y_{b_i}$, so we can replace $\mathrm{avg}(b_i, b_{b_i})$ with $y_{b_i}$. Moreover, instead of initializing $y_i$ with $\mathrm{avg}(i, b_i)$ in Line \ref{improved free model algorithm line update y}, we can first initialize $y_i$ with $z_i$, and update it when $b_i$ is updated, so that we can also replace $\mathrm{avg}(i, b_i)$ in Line \ref{improved free model algorithm line update b loop} with $y_i$. As a result, Line \ref{improved free model algorithm line update b loop} to \ref{improved free model algorithm line update y} can be rewritten as follows.
{
\setlength{\interspacetitleruled}{0pt}%
\setlength{\algotitleheightrule}{0pt}%
\begin{algorithm}[H]
\nonl$\cdots$\\
\nonl$y_i\leftarrow z_i$\;
 \nonl\While {$b_i\neq n+1$ and $y_i\le y_{b_i}$} {
      \nonl$\displaystyle y_i\leftarrow \frac{(b_i-i)y_i+\left(b_{b_i}-b_i\right)y_{b_i}}{b_{b_i}-i}$\;
      \nonl$b_i\leftarrow b_{b_i}$\; 
  }
 \nonl$\cdots$
\end{algorithm}
}

Now in this new pseudocode, there is no $\mathrm{avg}$ any more, so Algorithm \ref{improved free model algorithm} is able to run in $O(n\log n)$ time even if $\mathrm{avg}(i,j)$ cannot be computed in $O(1)$ time.

\begin{appendices}

\section*{Proof of Theorem \ref{prop theorem}}

\begin{proof}
If $x_1^{\text{prop}}=x_2^{\text{prop}}=0$, we have for any $0<\delta\le q_1$, $u_1(\delta,0)\le u_1(0,0)=0$, i.e., $\delta / q_1 \ge B/C$, which is impossible, so we can assume $x_1^{\text{prop}}$ and $x_2^{\text{prop}}$ cannot be 0 at the same time.

Without loss of generality, we assume $x_1^{\text{prop}}\neq 0$. Let $u_1(x)=u_1(x,x_2^{\text{prop}})$, then for any $0<x\le q_1$, $u_1'(x)=x_2^{\text{prop}}B/(x+x_2^{\text{prop}})-C/q_1$, so we have $u_1'(q_1)=x_2^{\text{prop}}B/(q_1+x_2^{\text{prop}})^2-C/q_1\le 0$. Since $x_1^{\text{prop}}$ is an optimal value for $u_1(x)$ over $(0,q_1]$, and $u_1'(q_1)\le 0$, we must have $u_1'(x_1^{\text{prop}})=0$, which means
\begin{equation}\label{x1prop}
x_2^{\text{prop}}=\frac{(x_1^{\text{prop}}+x_2^{\text{prop}})^2C}{q_1B}>0.
\end{equation}
Then after performing a similar argument as above, we have
\begin{equation}\label{x2prop}
x_1^{\text{prop}}=\frac{(x_1^{\text{prop}}+x_2^{\text{prop}})^2C}{q_2B}.
\end{equation}
By combining (\ref{x1prop}) and (\ref{x2prop}) we get $x_1^{\text{prop}}+x_2^{\text{prop}}=Bq_1q_2/(C(q_1+q_2))=q_1q_2/(q_1+q_2)$. So we can set $q_1=\min\{\epsilon,1/2\}$ and $q_2=1-q_1$, then $(x_1^{\text{prop}}+x_2^{\text{prop}})/q_2=q_1/(q_1+q_2)\le \epsilon$.

By contrast, if we set the reward function $f$ to be 
\[
f(x)=
\begin{cases}
0, &\text{if $x<q_2$,}\\
C, &\text{otherwise,}
\end{cases}
\]
then agent 1 would not participate while agent 2 is incentivized to produce a content with quality $q_2$. Hence we have $x_1^*+x_2^*=q_2$, which completes the proof.
\end{proof}

\section*{Proof of Lemma \ref{free model x lemma}}
\begin{proof}
Note we can arbitrarily reorder $x_i^*$'s for players with the same ability with keeping the solution feasible, we can assume $x_i^*\ge x_j^*$ if $q_i=q_j$ for each $i>j$ without loss of generality. We then show that with this assumption, $x_i^*\ge x_j^*$ for each $i>j$ even if $q_i>q_j$, which completes the proof. 

Let us consider indices $i>j$ with $q_i>q_j$. Suppose $x_j^*>x_i^*$. By the incentive constraint, we have $u_i(x_i^*)\ge u_i(x_j^*)$ and $u_j(x_j^*)\ge u_j(x_i^*)$, or $f(x_i^*)-x_i^*C/q_i\ge f(x_j^*)-x_j^*C/q_i$ and $f(x_j^*)-x_j^*C/q_j\ge f(x_i^*)-x_i^*C/q_j$. By summing up the two inequalities above, we have $(x_i^*-x_j^*)j(1/q_j-1/q_i)\ge0$, a contradiction. Hence $x_j^*\le x_i^*$.
\end{proof}

\section*{Proof of Lemma \ref{free model f lemma}}

\begin{proof}
Suppose $f$ with $x_1^*,x_2^*,\ldots,x_n^*$ is a feasible solution to (\ref{free problem}). Note by the incentive constraint, we have for any $i$, $u_i(x_i^*)\ge u_i(x_{i-1}^*)$, that is ($x_0^*$ is defined to be $0$)
\[f(x_i^*)-\frac{x_i^*C}{q_i}\ge f(x_{i-1}^*)-\frac{x_{i-1}^*C}{q_i}.\]
This inequality holds for any $i$, so we can sum up the inequalities for $1,2,\ldots,i$ to get
\[\sum_{j=1}^i f(x_j^*)-\sum_{j=1}^i \frac{x_j^*C}{q_j}\ge \sum_{j=1}^i  f(x_{j-1}^*)-\sum_{j=1}^i \frac{x_{j-1}^*C}{q_j},\]
i.e.,
\begin{equation} \label{f(x_i^*) inequality}
f(x_i^*)\ge \left(\frac{x_i^*}{q_i}+\sum_{j=1}^{i-1}\left(\frac{1}{q_j}-\frac{1}{q_{j+1}}\right)x_j^*\right)C.
\end{equation}
By summing up (\ref{f(x_i^*) inequality}) from $i=1$ to $n$, we have
\[\sum_{i=1}^nf(x_i^*)\ge C\left(\frac{x_n^*}{q_n}+\sum_{i=1}^{n-1}\left((n-i)\left(\frac{1}{q_i}-\frac{1}{q_{i+1}}\right)+\frac{1}{q_i}\right)x_i^*\right).\]
Since by the budget constraint, $\sum_{i=1}^nf(x_i^*)\le B$, the inequality (\ref{free model budget constraint}) is proven.

On the other hand, suppose (\ref{free model budget constraint}) is satisfied, we can choose $f$ to be a step function according to (\ref{f(x_i^*) inequality}), i.e.,
\begin{equation}
f(x)=\begin{dcases}
0, & \text{if } 0\le x<x_1^*, \\
\dfrac{x_1^*C}{q_1}, & \text{if } x_1^*\le x<x_2^*, \\
\left(\dfrac{x_1^*}{q_1}-\dfrac{x_1^*}{q_2}+\dfrac{x_2^*}{q_2}\right)C, & \text{if } x_2^*\le x<x_3^*, \\
\cdots, \\
\left(\dfrac{x_1^*}{q_1}-\dfrac{x_1^*}{q_2}+\dfrac{x_2^*}{q_2}-\dfrac{x_2^*}{q_3}+\cdots+\dfrac{x_n^*}{q_n}\right)C, & \text{if } x\ge x_n^*.
\end{dcases}
\end{equation}
We can check that $f$ with $x_1^*,x_2^*,\ldots,x_n^*$ is a feasible solution to (\ref{free problem}).
\end{proof}

\section*{Proof of Lemma \ref{B lemma}}

\begin{proof}
Consider the last iteration, i.e., the iteration where $i^*=i_k$. If $d=\hat{B}/((i_R-i^*)y_i)$ in Line \ref{free model algorithm line compute d}, then $\hat{B}$ is updated to 0 in Line \ref{free model algorithm line update B}. Now suppose $\sum_{i=1}^n z_ix_i=K$, then $\hat{B}\neq 0$ by (\ref{B}), thus $d$ must be $q_{i^*}-x_{i^*}$, therefore $x_i^*$ is updated to $q_{i^*}$ in Line \ref{free model algorithm line update x}, hence $x_{i_k}=q_{i_k}$ after this iteration. Moreover, since $\hat{B}\neq 0$ but the algorithm halts after this iteration, so we must have $S=\{0,1,\ldots,n+1\}$ by the condition in Line \ref{free model algorithm line loop begin}. This means immediately after the $(\ell-1)$th iteration, $S=\{0,1,\ldots,n+1\}-i_k$, and by (\ref{x}), for each $i\neq i_k$, $x_i=q_i$. Hence $x_i=q_i$ for each $i$.
\end{proof}

\section*{Proof of Lemma \ref{free model algorithm correctness lemma}}

\begin{proof}
We prove there exists an optimal solution for Problem (\ref{main form}) such that for any $\ell \le k$, $x_{i_1}^*=x_{i_1},x_{i_2}^*=x_{i_2},\ldots$, and $x_{i_\ell}^*=x_{i_\ell}$, by mathematical induction on $\ell$. 

The case where $\ell=0$ is trivial.

Now suppose there exists an optimal solution for Problem (\ref{main form}) such that $x_{i_1}^*=x_{i_1},x_{i_2}^*=x_{i_2},\ldots$, and $x_{i_{\ell-1}}^*=x_{i_{\ell-1}}$, then we have $x_{i_1}^*=x_{i_1}=q_{i_1},x_{i_2}^*=x_{i_2}=q_{i_2},\ldots$, and $x_{i_{\ell-1}}^*=x_{i_{\ell-1}}=q_{i_{\ell-1}}$ by (\ref{x}). We call an index $i$ \emph{irregular} if $i\notin S_\ell$ and $x_i^*>x_{i-1}^*$ ($x_0^*$ is defined to be $0$). 

Now there are several cases.

\begin{enumerate}
\item\label{x*=x case} If $x_{i_{\ell}}^*=x_{i_{\ell}}$, the proof is completed. 

\item If $x_{i_{\ell}}^*>x_{i_{\ell}}$, then in this case, $x_{i_1}^*=x_{i_1},x_{i_2}^*=x_{i_2},\ldots, x_{i_\ell}^*>x_{i_\ell}$, and for any $i\notin S_\ell$, 
\begin{align*}
x_i^*&\ge x_{\mathrm{left}_\ell(i)}^* \tag*{(by the feasibility constraints in Problem (\ref{main form}))}\\
&=x_{\mathrm{left}_\ell(i)}  \tag*{(by the inductive assumption)}\\
&=x_i.  \tag*{(by (\ref{x}))}
\end{align*}
So we have $\sum_{i=1}^n z_ix_i^*>\sum_{i=1}^n z_ix_i$. Also, since $x_{i_{\ell}}<x_{i_{\ell}}^*\le q_{i_\ell}$, by Lemma \ref{B lemma} we have $\sum_{i=1}^n z_ix_i=K$, so $\sum_{i=1}^n z_ix_i^*>K$, which contradicts to the feasibility constraints in Problem (\ref{main form}). 

\item\label{no irregular case} If $x_{i_{\ell}}^*<x_{i_{\ell}}$ and there is no irregular index, then consider the point immediately after the $\ell$th iteration. At this point, we have if $i\notin[i_\ell,\mathrm{right}_\ell(i_\ell))$,
\begin{align*}
x_i^*&= x_{\mathrm{left}_\ell(i)}^* \tag*{(since there is no irregular index)}\\
&=x_{\mathrm{left}_\ell(i)}  \tag*{(by the inductive assumption)}\\
&=x_i,  \tag*{(by (\ref{x}))}
\end{align*}
and otherwise,
\begin{align*}
x_i^*&= x_{i_\ell}^* \tag*{(since there is no irregular index)}\\
&<x_{i_\ell}  \tag*{(by the assumption of Case \ref{no irregular case})}\\
&=x_i,  \tag*{(by (\ref{x}))}
\end{align*}
thus $\sum_{i=1}^n z_ix_i>\sum_{i=1}^n z_ix_i^*$. Also, by (\ref{B}) we have $K-\sum_{i=1}^nz_ix_i=\hat{B}\ge 0$, so $K-\sum_{i=1}^nz_ix_i^*>K-\sum_{i=1}^nz_ix_i\ge 0$. As a result, we can increase $x_{i_{\ell}}^*,x_{i_{\ell}+1}^*,\ldots,x_{\mathrm{right}_\ell(i_{\ell})-1}^*$ a bit with keeping $\sum_{i=1}^nz_ix_i^*\le K$ and increasing $\sum_{i=1}^n x_i^*$, which contradicts to the fact that $x_1^*,x_2^*,\ldots,x_n^*$ make an optimal solution for Problem (\ref{main form}). 

\item \label{irregular in case} If $x_{i_{\ell}}^*<x_{i_{\ell}}$ and there exists an irregular index in $(i_\ell,\mathrm{right}_\ell(i_\ell))$, let $r$ be the smallest such index, and define
\begin{align*}
M=\min\{ (x_r^*-x_{r-1}^*)\mathrm{sum}(r, \mathrm{right}_\ell(r)), (q_{i_\ell}-x_{i_\ell}^*)\mathrm{sum}(i_\ell, \mathrm{right}_\ell(i_\ell)) \}.
\end{align*}
Consider an operation that for any $r\le i<\mathrm{right}_\ell(r)$, decreases $x_i^*$ by $M/\mathrm{sum}(r, \mathrm{right}_\ell(r))$, then for any $i_{\ell}\le i<\mathrm{right}_\ell(i_\ell)$, increases $x_i^*$ by $M/\mathrm{sum}(i_\ell, \mathrm{right}_\ell(i_\ell))$. This operation is diagramed in Figure \ref{before figure} and \ref{after figure}, and has the following properties:
\begin{figure}
\centering
\begin{minipage}{0.45\textwidth}
\centering
\begin{tikzpicture}[scale=0.7]
    \begin{axis}[
        symbolic x coords={1, 2, 3, 4, 5, 6, 7},
        ylabel=$x_i^*$,
        ylabel near ticks, 
        ylabel style={rotate=-90},
        ybar, 
        ymin=0,
        ymax=3.2,
        bar width=20pt, 
        yticklabels={},
        xticklabels={$\cdots$\vphantom{1}, $i_\ell$, $i_\ell+1$, $\cdots$\vphantom{1}, $r$\vphantom{1}, $r+1$, $\cdots$\vphantom{1}},
        xtick=data
    ]
    \addplot[pattern=north east lines] coordinates {
        (1, 0)
        (2,   1)
        (3,  1)
        (4, 0)
        (5,  2)
        (6,  3)
        (7, 0)
    };
    \end{axis}
\end{tikzpicture}
\caption{Before the Operation}\label{before figure}
\end{minipage}\hfill
\begin{minipage}{0.45\textwidth}
\centering
\begin{tikzpicture}[scale=0.7]
    \begin{axis}[
        symbolic x coords={1, 2, 3, 4, 5, 6, 7},
        ylabel=$x_i^*$,
        ylabel near ticks, 
        ylabel style={rotate=-90},
        ybar, 
        ymin=0,
        ymax=3.2,
        bar width=20pt, 
        yticklabels={},
        xticklabels={$\cdots$\vphantom{1}, $i_\ell$, $i_\ell+1$, $\cdots$\vphantom{1}, $r$\vphantom{1}, $r+1$, $\cdots$\vphantom{1}},
        xtick=data
    ]
    \addplot[pattern=north east lines] coordinates {
        (1, 0)
        (2,   1.5)
        (3,  1.5)
        (4, 0)
        (5,  1.5)
        (6,  2.5)
        (7, 0)
    };
    \end{axis}
\end{tikzpicture}
\caption{After the Operation}\label{after figure}
\end{minipage}
\end{figure}

\begin{enumerate}
\item It does not decrease $\sum_{i=1}^nx_i^*$.

Because in the $\ell$th iteration, the algorithm chooses $i_\ell$ instead of $r$ in the $\ell$th iteration, by (\ref{y}) we have
$\mathrm{avg}(i_\ell, \mathrm{right}_\ell(i_\ell))=y_{i_\ell}\le y_r=\mathrm{avg}(r, \mathrm{right}_\ell(r))$. So this operation increases $\sum_{i=1}^nx_i^*$ by
\begin{align*}
&-\frac{(\mathrm{right}_\ell(r)-r)M}{\mathrm{sum}(r, \mathrm{right}_\ell(r))}+\frac{(\mathrm{right}_\ell(i_\ell)-i_\ell)M}{\mathrm{sum}(i_\ell, \mathrm{right}_\ell(i_\ell))}\\
\ &=-\frac{M}{\mathrm{avg}(r, \mathrm{right}_\ell(r))}+\frac{M}{\mathrm{avg}(i_\ell, \mathrm{right}_\ell(i_\ell))}\\
\ &\ge 0.
\end{align*}
\item It keeps the solution feasible. 
\begin{enumerate}
\item The $x_{i-1}^*\le x_i^*\le q_i$ constraint is still satisfied.

Note $i_\ell <r<\mathrm{right}_\ell(i_\ell)$ due to the definition of $r$, we have $\mathrm{right}_\ell(r)=\mathrm{right}_\ell(i_\ell)$ due to the definition of $r$, and $\mathrm{sum}(r, \mathrm{right}_\ell(r))\le\mathrm{sum}(i_\ell, \mathrm{right}_\ell(i_\ell))$, so $x_r^*, x_{r+1}^*,\ldots,x_{\mathrm{right}_\ell(r)-1}^*$ are not increased, and are decreased by at most
\[\frac{M}{\mathrm{sum}(r, \mathrm{right}_\ell(r))}\le\frac{(x_r^*-x_{r-1}^*)\mathrm{sum}(r, \mathrm{right}_\ell(r))}{\mathrm{sum}(r, \mathrm{right}_\ell(r))}=x_r^*-x_{r-1}^*,\]
which means the $x_{i-1}^*\le x_i^*\le q_i$ constraint is still satisfied for $x_r^*, x_{r+1}^*,\ldots,x_{\mathrm{right}_\ell(r)-1}^*$. Also, $x_{i_\ell}^*=x_{i_\ell+1}^*=\cdots=x_{r-1}^*$ since $i_\ell,i_\ell+1,\ldots, r-1$ are not irregular, and they are increased by at most 
\[\frac{M}{\mathrm{sum}(i_\ell, \mathrm{right}_\ell(i_\ell))}\le\frac{(q_{i_\ell}-x_{i_\ell}^*)\mathrm{sum}(i_\ell, \mathrm{right}_\ell(i_\ell))}{\mathrm{sum}(i_\ell, \mathrm{right}_\ell(i_\ell))}=q_{i_\ell}-x_{i_\ell}^*,\]
so $x_{i_\ell}^*,x_{i_\ell+1}^*,\ldots,x_{r-1}^*$ also satisfy this constraint.

\item The $\sum_{i=1}^n z_ix_i^*\le K$ constraint is still satisfied.

Because $\sum_{i=1}^n z_ix_i^*$ is increased by exactly
\begin{align*}
-\frac{\displaystyle M\mathrm{sum}(r, \mathrm{right}_\ell(r))}{\mathrm{sum}(r, \mathrm{right}_\ell(r))}
+\frac{\displaystyle M\mathrm{sum}(i_\ell, \mathrm{right}_\ell(i_\ell))}{\mathrm{sum}(i_\ell, \mathrm{right}_\ell(i_\ell))}= 0.
\end{align*}
\end{enumerate}
\item It keeps $x_{i_1}^*=x_{i_1},x_{i_2}^*=x_{i_2},\ldots$, and $x_{i_{\ell-1}}^*=x_{i_{\ell-1}}$ because they are not updated by this operation.
\item It either decreases the number of irregular indices, or make $x_{i_\ell}^*=x_{i_\ell}$.

If $M=(x_r^*-x_{r-1}^*)\mathrm{sum}(r, \mathrm{right}_\ell(r))$, then $x_r^*, x_{r+1}^*,\ldots,x_{\mathrm{right}_\ell(r)-1}^*$ are first decreased by exactly $x_r^*-x_{r-1}^*$, which makes $r$ not irregular and does not make any non-irregular index irregular. Then $x_{i_\ell},x_{i_\ell+1},\ldots,x_{\mathrm{right}_\ell(i_\ell)-1}$ are increased, which does not make any non-irregular index irregular. 

On the other hand, if $M=(q_{i_\ell}-x_{i_\ell}^*)\mathrm{sum}(i_\ell, \mathrm{right}_\ell(i_\ell)$, then $x_{i_\ell}^*$ is increased by exactly $q_{i_\ell}-x_{i_\ell}^*$, so $x_{i_\ell}^*=x_{i_\ell}$ after this operation.
\end{enumerate}
So we can apply this operation again and again with keeping the solution feasible and optimal until $x_{i_\ell}^*=x_{i_\ell}$ or there is no irregular index in $(i_\ell,\mathrm{right}_\ell(i_\ell))$, which falls into Case \ref{x*=x case} or Case \ref{no irregular case} or \ref{irregular beyond case} respectively.

\item \label{irregular beyond case} If $x_{i_{\ell}}^*<x_{i_{\ell}}$ and all irregular indices (at least one) do not belong to $(i_\ell,\mathrm{right}_\ell(i_\ell))$, let $r$ be the smallest irregular index, then we can define $M$ and the operation in the same way as Case \ref{irregular in case}. The properties of this operation in Case \ref{irregular in case} still hold. Properties (a), (b)(ii), (c) and (d) hold due to exactly the same argument. Property (b)(i) holds because $x_r^*,x_{r+1}^*,\ldots,x_{\mathrm{right(r)-1}}^*$ are decreased by at most $x_r^*-x_{r-1}^*$, and $x_{i_\ell}=x_{i_\ell+1}=\cdots=x_{\mathrm{right}_\ell(i_\ell)-1}$ and they are increased by at most $q_{i_\ell}-x_{i_\ell}^*$. Hence, we can apply this operation again and again until $x_{i_\ell}^*=x_{i_\ell}$ or there is no irregular index any more, which falls into Case \ref{x*=x case} or Case \ref{no irregular case} respectively.
\end{enumerate}
\end{proof}

\section*{Proof of Lemma \ref{yi lemma}}

\begin{proof}
Note in each iteration, for any $i_L<i<i^*$, Algorithm \ref{free model algorithm} chooses $i^*$ instead of $i$, so $y_i\ge y_{i^*}$. By (\ref{y}), we have $\mathrm{avg}(i,i_R)=y_i\ge y_{i^*}=\mathrm{avg}(i^*,i_R)$, so $\mathrm{avg}(i,i_R)\le\mathrm{avg}(i,i^*)$, or $y_i\le\mathrm{avg}(i,i^*)$, which means each update does not decrease $y_i$. So in each iteration, we have $y_i\le \mathrm{avg}(i,b(i))$, and $y_{i^*}=\mathrm{avg}(i^*,b(i^*))$ since $y_{i^*}$ will not be updated any more. Hence $i^*=\arg\min_{i\in S} (y_i,i)=\arg\min_{i\in S} (\mathrm{avg}(i,b(i)), i)$, which means the modified algorithm will choose the same $i^*$ in each iteration, and Line \ref{free model algorithm line loop begin} to \ref{free model algorithm line update x} have the same effect, so the modified algorithm would output the same $x_1,x_2,\ldots,x_n$ as Algorithm \ref{free model algorithm}.
\end{proof}

\section*{Proof of Lemma \ref{iRb lemma}}

\begin{proof}
Note in each iteration, $y_i$ is updated if and only if $i_L< i<i^*$. Also, if $i\notin S$ at the beginning of this iteration, $\min_{j\in S:j\ge i}j$ is updated in this iteration if and only if $i_L< i<i^*$. Furthermore, $i^*$ is exactly the new $\min_{j\in S:j>i}j$. Hence, before $i$ is added to $S$, $\min_{j\in S:j>i}j$ is updated if and only if $y_i$ is updated in the same iteration, and since $y_i$ will not be updated any more after $i$ is added to $S$, $b(i)$ records the last updated $\min_{j\in S:j>i}j$ before $i$ is added to $S$, which means $b(i^*)=i_R$ in each iteration.
\end{proof}

\section*{Proof of Lemma \ref{order lemma}}

\begin{proof}
Consider the iteration where $i^*=i$. By Lemma \ref{iRb lemma}, $b(i)$ is already in $S$ while $j$ is not for the set $S$ immediately before this iteration. Note an index is added to $S$ exactly during the iteration where $i^*$ is chosen to be this index, so the iteration where $i^*=b(i)$ comes before the iteration where $i^*=i$, which comes before the iteration where $i^*=j$.
\end{proof}

\section*{Proof of Lemma \ref{avg lemma}}

\begin{proof}
For any $i^*<i<i_R$, we have $\mathrm{avg}(i^*,b(i^*))=y_{i^*}\le y_i=\mathrm{avg}(i,b(i^*))$, so $\mathrm{avg}(i^*,i)\le \mathrm{avg}(i,b(i^*))$. For any $i_L<i<i^*$, we have $\mathrm{avg}(i^*,b(i^*))=y_{i^*}< y_i=\mathrm{avg}(i,b(i^*))$ (note if $y_{i^*}< y_i$ in this case, the index $i$ would be chosen instead of $i^*$ by Line \ref{free model algorithm line loop begin} of Algorithm \ref{free model algorithm}, so we use $<$ instead of $\le$ here), so $\mathrm{avg}(i,b(i^*))<\mathrm{avg}(i, i^*)$.
\end{proof}

\section*{Proof of Lemma \ref{iL lemma}}

\begin{proof}
By Lemma \ref{order lemma}, $b^m(j), b^{m-1}(j),\cdots,b^0(j)=j$ are added to $S$ in order. For any $j\le k <i$, suppose $b^t(j)\le k<b^{t+1}(j)$. Consider the iteration where $i^*=b^t(j)$. By Lemma \ref{iRb lemma}, $k$ does not belong to $S$ at the beginning of the iteration where $i^*=b^t(j)$, so $k$ does not belong to $S$ at the beginning of the iteration where $i^*=b^m(j)$, since it is an earlier iteration. Note this holds for all $j\le k <i$, so $i_L<j$ in the iteration where $i^*=b^m(j)$.
\end{proof}
\end{appendices}

\bibliographystyle{unsrtnat}  
\bibliography{ref}

\end{document}